\documentclass[prl,a4paper,superscriptaddress,twocolumn,showpacs,amsmath,amssymb,floatfix]{revtex4}

\usepackage{graphicx}
\usepackage{amssymb}
\usepackage{float}
\usepackage{hyperref}

\usepackage{color}

\begin{document}

\title{Boltzmann-Loschmidt dispute reloaded quantum 150 years later}

\author{Leonardo Ermann}
\affiliation{{\mbox DFTIFSC, GIyA, CNEA,  Av. Libertador 8250, C1429 CABA, Argentina}}
\affiliation{\mbox{Consejo Nacional de Investigaciones Cient\'{\i}ficas y T\'ecnicas (CONICET),
    C1425FQB CABA, Argentina}}
\author{Alexei D. Chepelianskii}
\affiliation{\mbox {LPS, Universit\'e Paris-Sud, CNRS, UMR 8502, Orsay F-91405, France}}
\author{Dima L. Shepelyansky}
%\homepage[]{http://www.quantware.ups-tlse.fr}
\affiliation{\mbox{ 
Univ Toulouse, CNRS, Laboratoire de Physique Th\'eorique,  Toulouse, France}}

%\date{today}
\date{April 3, 2026}

\begin{abstract}
  The Boltzmann-Loschmidt dispute of 1876 questioned the
  possibility of a statistical irreversible description by time
  reversible classical equations of motion of atoms.
  Here we show analytically and numerically that the quantum chaos diffusion
  of cold atoms, or ions, in a harmonic trap and pulsed optical lattice
  can be inverted back in time with up to 100\% efficiency.
  This is in sharp contrast to classical evolution where
  exponentially small errors break time reversibility.
  We argue that the existing experimental skills allow highlighting
  the Boltzmann-Loschmidt dispute from a  quantum perspective.
\end{abstract}

%\pacs{05.45.Mt, 67.85.Hj,  47.27.-i, 72.15.Rn}
%05.45.-a Nonlinear dynamics and chaos
%67.85.Hj 	Bose-Einstein condensates in optical potentials
%47.27.-i 	Turbulent flows
%72.15.Rn 	Localization effects (Anderson or weak localization) 
%
%47.35.-i 	Hydrodynamic waves
%47.35.Bb 	Gravity waves 
%89.75.-k 	Complex systems 

\maketitle

{\it Introduction.-} In 1872 Boltzmann formulated the statistical theory
of entropy growth  and thermalization  based
 on the dynamical laws of classical motion
 of atoms  \cite{boltzmann1}.
 A few years later in 1876, 150 years ago,  this theory
 was objected by Loschmidt \cite{loschmidt} who pointed
 that the dynamics of atoms
 is reversible in time thus raising a question
 of how an irreversible thermalization can
 appear from the reversible dynamical equations of motion.
 The reply of Boltzmann  followed in 1877 \cite{boltzmann2}
 and the legend tells that on the direct question
 of what happens with entropy and thermalization
 if one inverts velocities of all atoms
 he replied {\it then go and invert them} \cite{mayer}.
 The modern resolution of this Boltzmann-Loschmidt dispute
 is given by the theory of dynamical chaos
 in generic nonlinear systems with  positive Lyapunov
 exponent $\Lambda$ and Kolmogorov-Sinai entropy $h_{KS}$ 
 leading to an exponential instability of motion
 \cite{arnold,sinai,chirikov1979,lichtenberg}.
 This instability generates  an   exponential growth of
 errors thus breaking time reversibility
 even if time reversal errors are negligibly small
 (see e.g. an example in \cite{dls1983}).
 It also leads to mixing
 with time at exponentially smaller and smaller
 scales in the phase space.

 We should point that 
 the time reversibility problem of statistical laws,
 originated by the Boltzmann-Loschmidt dispute
 \cite{boltzmann1,loschmidt}, is still
 actively discussed by a scientific community
 in  physics and philosophy
 (see e.g. \cite{sklar,lindey,bader,price,uffink,uffink1,weidenmuller}).

However, the above discussions are mainly done in the frame of classical mechanics
while the reality is quantum. The quantum evolution
is generally described by the linear Schr\"odinger
equation and a chaotic mixing in a phase space
stops at the Planck constant $\hbar$ being protected by
the Heisenberg uncertainty relation.
Indeed, due to exponential divergence of classical trajectories
the Ehrenfest theorem for  wave packet spreading \cite{ehrenfest}
remains valid only for a logarithmically short Ehrenfest time
$t_E \sim |\ln \hbar |/\Lambda $ so that after $t_E$
the wave packet spreads exponentially  over the main part of phase space
\cite{chirikov1981,chirikov1986,haake}
and there is no instability for times $t>t_E$
\cite{dls1981,dlsehrenf}. Thus
the time reversibility is preserved for the quantum evolution
(see an example in \cite{dls1983}).
The properties of time reversibility in systems of quantum chaos \cite{haake}
have been studied in detail being known
as Loschmidt echo and fidelity decay
(see e,g, \cite{peres,jalabert1,frahm,prosen,jacquod,jalabert2} and Refs. therein).

The first experiments on time reversal  had been done with spin echos \cite{hahn1,hahn2,pastaw}.
Later the time reversal was realized with acoustic 
and electromagnetic waves that led to important useful applications
including seismic analysis in geophysics (see e.g. \cite{fink1,fink2,geopht}).
However, in far 1876 Boltzmann and Loschmidt
discussed the time reversibility of atoms
and for this system an experimental realization
of time reversal of atomic matter waves 
is rather nontrivial. A possible realization
of time reversal of a quantum chaos evolution of  cold atoms in a kicked optical
lattice was proposed in \cite{qbl1}
and for a Bose-Einstein condensate (BEC) of atoms in \cite{qbl2}.
The main elements of this proposal are based on a possibility to
transfer amplitude $k$ of kicks from positive to negative
value and the property of free propagation of atoms
between kicks with a phase factor
$U \sim \exp(-iT (n+\alpha)^2)/2)$ where a parameter  $T$ is
proportional to a period between kicks
and a fractional part $\alpha$ of momentum of atom $p=n+\alpha$
is not affected by kicks due to periodicity of
optical lattice. Due to that a time reversal
is realized by changing $T=4\pi+\varepsilon$ to $T=4\pi - \varepsilon$
and by $k \rightarrow -k$ at the middle of time interval between kicks.
However, this time reversal is exact only for atoms
with a fractional quasimomentum $\alpha=0$
while for $\alpha$ close to zero
the time reversal degradates with an increase of time
interval $t_r$ at which the time reversal operation is performed.
Thus time reversal works only for a relatively small
group of atoms with $\alpha \approx 0$. The interactions
 between atoms also lead to a decrease of return signal \cite{qbl2}.
The time reversal of atomic matter waves was realized in
BEC experiments \cite{bec1,bec2}. However, due to
an approximate nature of time reversal
 for atoms with $\alpha \approx 0$ only small
values of time interval $t_r=5$ were realized
with 5 kicks of forward and backward propagation.

\begin{figure}[t]
\begin{center}
\includegraphics[width=0.48\textwidth]{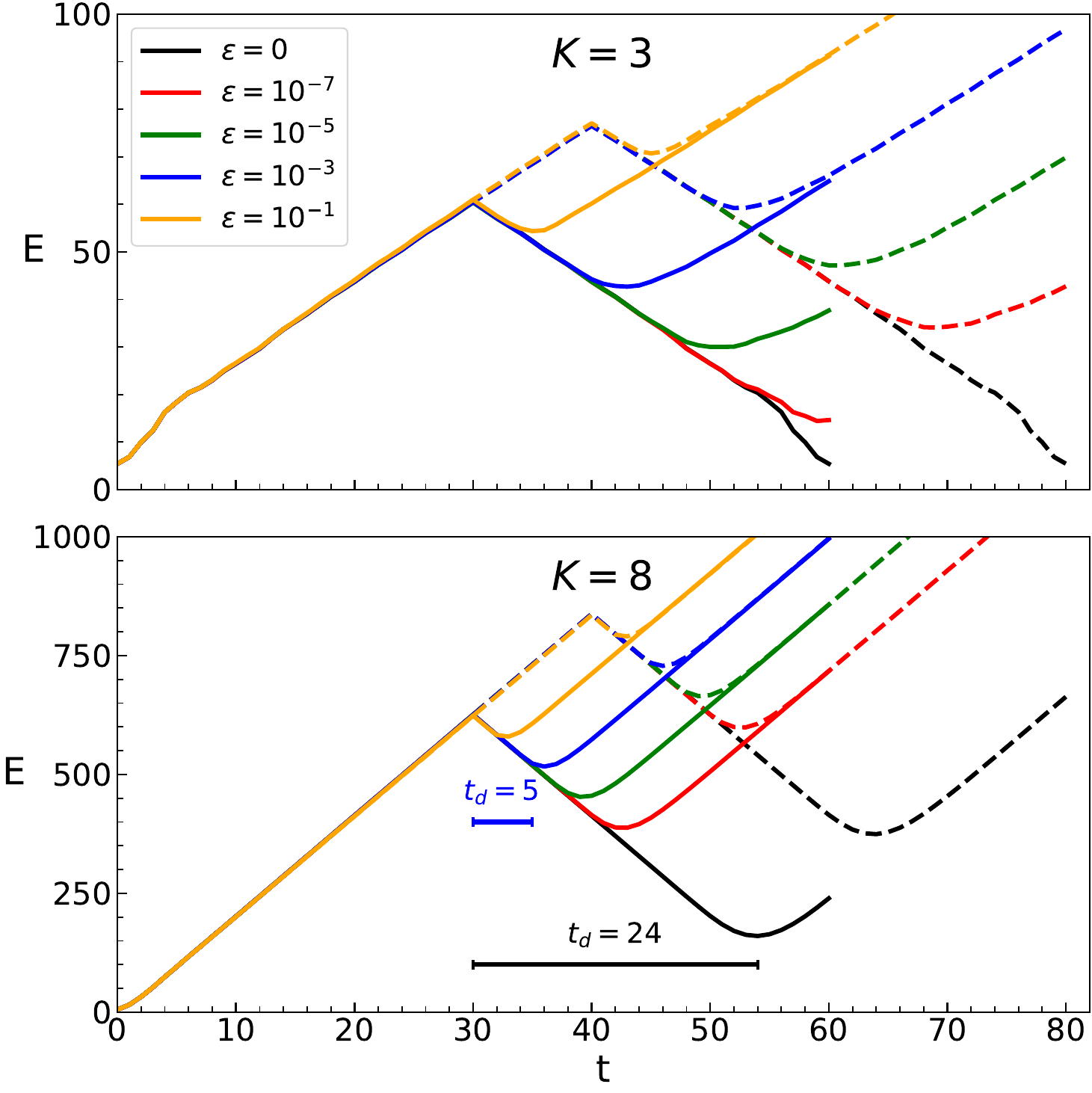}
\end{center}
\vglue -0.3cm
\caption{\label{fig1}
  Time dependence of the mean energy $E(t)$ for the classical system 
  (\ref{eq1}) at $K=3$ (top panel) and $K=8$ (bottom panel). 
  Time reversal is performed at $t_r=30$ (solid curves) and 
  $t_r=40$ (dashed curves) in the presence of noise with 
  amplitude $\varepsilon$. 
  Values of $E$ are averaged over $10^6$ classical trajectories. 
  At $t=0$, the initial distribution is a Gaussian centered at the unstable 
  fixed point $(x_0=\pi, p_0=0)$ with a standard deviation 
  $\sigma = \sqrt{2}/2$. 
  The recovery time $t_d$ is defined as the time during which the energy 
  decays after time reversal, as illustrated in the bottom panel. 
  For $\varepsilon=0$, no artificial noise is added, 
  leaving only computer round-off errors at 
  the double-precision level ($\sim 10^{-16}$).
}
\end{figure}

{\it Model description.-} In this work we show that the time reversal
can be realized with cold atoms
placed in a harmonic trap and kicked optical lattice
and that in this system of quantum chaos
almost all atoms return to the origin
after a long reversal time $t_r$ 
with probability close to 100\%. The system is described
by the Hamiltonian
\begin{equation}
  \hat{H} = (\hat{p}^2+ {\omega_0}^2\hat{x}^2)/2  + 
  K \cos(q \hat{x})\sum_m \delta(t - m T) \; .
\label{eq1}
\end{equation}
where $\omega_0$ is a frequency of harmonic trap,
momentum $\hat{p}$ and position $\hat{x}$ operators have
the usual commutator $[\hat{p},\hat{x}] = -i \hbar$,
  $T$ is a time interval between kicks of optical lattice
  with a potential $V(x) = K \cos(qx)$ with
  space period $2\pi/q$. The classical dynamics of this system
  is described by the Hamiltonian  equations
  while the quantum evolution is described by the Schr\"odinger
  equation with the Planck constant $\hbar$. Between kicks
  we have evolution in a harmonic potential
  and a kick transfers a wave function $\psi$ to ${\bar \psi}$
  as ${\bar \psi} = \exp[-i(K/\hbar) \cos(qx)] \psi$.
  In our studies we use dimensionless units
  with atom mass and frequency $\omega_0$ being unity,
  $K, \hbar, T$ being dimensionless and $q=1$
  (the case of $q \neq 1$ is reduced to
  $q=1$ by a rescaling $K/\hbar  \rightarrow K/\hbar_{eff} = K/(\hbar q^2)$).

The classical system  (\ref{eq1}) was introduced and studied in
\cite{zaslav1,zaslav2,zaslavweb} being known as Zaslavsky web map.
This simple symplectic map describes a change 
of $p,x$ variables after one period of time $T$.
The dynamics depends on the ratio of oscillator period to the time between
kicks $R=2\pi/T$. For $R=3,4,6$ the separatrix web
covers the whole phase space plane $(x,p)$
corresponding to the known result of a plane covered
by triangles, squares and hexagons.
The Kolmogorov-Arnold-Moser (KAM) theory \cite{arnold,sinai,chirikov1979,lichtenberg}
is not applicable for such a case and even
at small $K$ values there are chaotic layers around separatrix lines
of a width proportional to $K$. For large $K$ values
the whole phase space is chaotic without visible stability islands
and the system energy $E= <(p^2+x^2)/2>$ is growing diffusely
with the number of kicks denoted as $t$ in the following
($E \approx Dt, D \approx K^2/4 $).
The map on one period is:
${\bar x} = p+ K\sin x, \;  {\bar p} = -x$,
where bar marks new values of variables
and $R=4, q=1$.
A variety of images of classical dynamics in the phase space
is available at \cite{webspace}.

The quantum evolution of  system  (\ref{eq1})
is described by the quantum map for the wave function
after one period of perturbation:
\begin{equation}
  {\bar \psi} = \exp(-i (\omega_0 T \hat{n} + \varepsilon_n))
  \exp[-i(K/\hbar) \cos q\hat{x} ]\psi \; ,
\label{eq2}
\end{equation}
where $\hat{n} = \hat{a}^+ \hat{a}$
is the standard operator of oscillator quantum number $n$
and we assume that certain experimental imperfections
at each map iteration induce random phases
$-\varepsilon_q \leq  \varepsilon_n \leq \varepsilon_q$
at oscillator levels;
in the following we use $\omega_0=1, \hbar=q=1$.
This quantum model at $\varepsilon_q=0$ was studied by different groups
(see e.g. \cite{sire,dana,zoller,kells,buchleitner,gardiner}
and Refs. therein). Quantum interference may lead
to localization of classical diffusion,
similar to a case of free cold atoms in
a kicked optical lattice
(see \cite{chirikov1981,chirikov1986,raizen}),
but there are also cases when the diffusion in energy is unlimited.
Here we consider the case of $R=4$ 
with a duality between coordinate and momentum
when the system (\ref{eq1}) can be reduced to the kicked Harper model
with unlimited quantum diffusion (see e.g. \cite{lima,sire,artuso}).
Numerically it is convenient to perform the quantum evolution
(\ref{eq2}) in the basis of oscillator eigenfunctions
using the matrix elements of  kick function
between these eigenstates (see e.g. \cite{flux}
where the quantum evolution (\ref{eq1})
was studied in presence of dissipation).

The time reversal of classical dynamics is done
by inversion of  velocities of all particles
$p \rightarrow -p$ at the middle of free rotation
between kicks. For the quantum evolution
one cannot perform experimentally the complex conjugation 
$\psi \rightarrow \psi^+$ but it is possible to
invert evolution backwards in time
by changing $T=2\pi/R$ to $T'=2\pi - T$
and replacing $K$ amplitude by $-K$
(at the moment of time reversal one should omit
one kick replaced by $T'$ rotation
and then followed by kicks with $-K$ amplitude
and rotation periods $T'$).
This time reversal operation
works also for irrational $R$ values.
Also such a time reversal
can be done if we add any integer
number multiplied by $2\pi$ to $T$ and $T'$.

{\it Time reversal of classical chaos.-}
The results for time reversal of classical dynamics
of Zaslavsky web map with Hamiltonian (\ref{eq1})
are shown in Fig.~\ref{fig1} for energy
time dependence $E(t)$. Energy is averaged over
$N=10^6$ trajectories with a Gaussian initial distribution 
centered at $x_0=\pi, p_0=0$ and standard deviation $\sigma=\sqrt{2}/2$ 
in the phase space. Due to chaos
there is a diffusive energy
growth with time $E = D t$
at the diffusion rate $D \approx K^2/4$
corresponding to random phase approximation
(actual values are $D/K^2 \approx 0.16 \ (K=3)$ and $0.33 \ (K=8)$
 due to presence of residual phase correlations, see \cite{chirikov1979,lichtenberg}).

\begin{figure}[t]
\begin{center}
\includegraphics[width=0.48\textwidth]{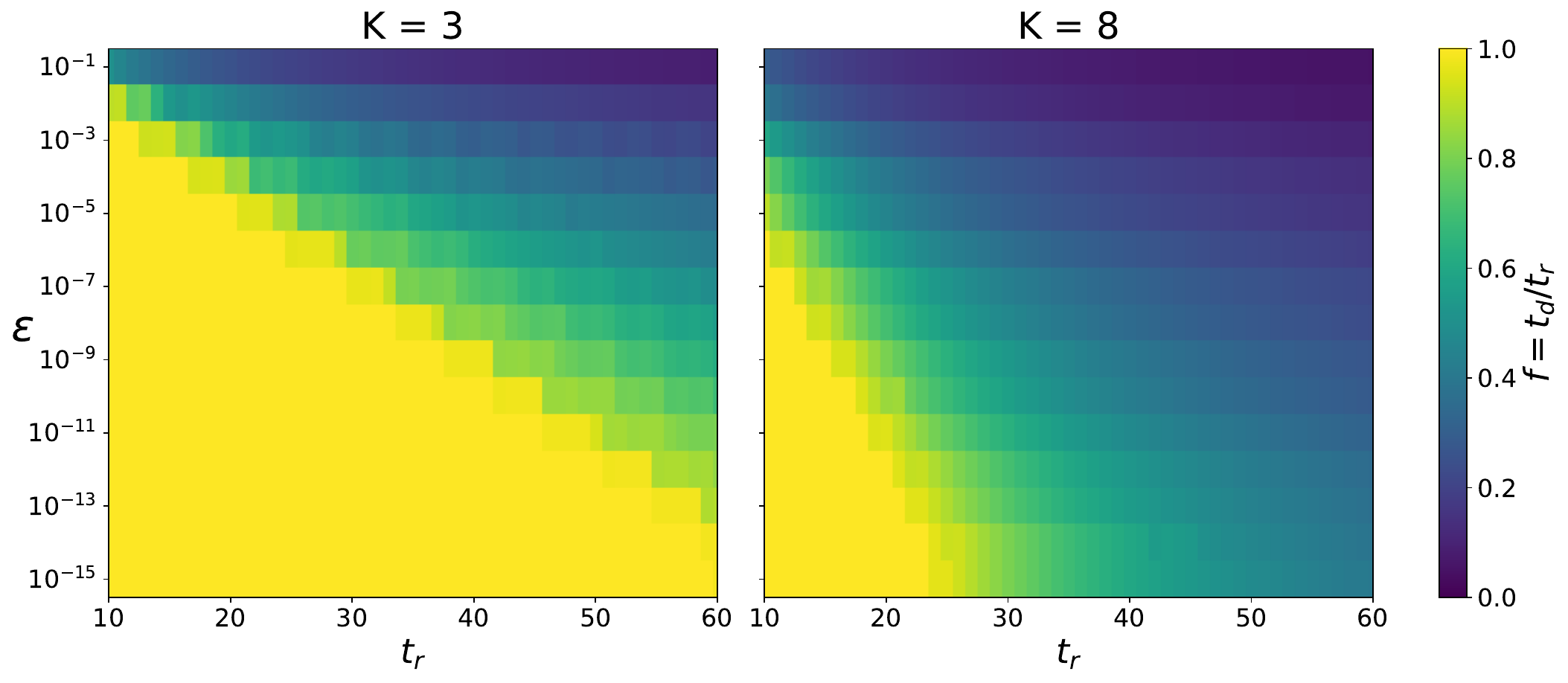}
\end{center}
\vglue -0.3cm
\caption{\label{fig2}
  Dependence of the ratio  of recovery and 
  reversal times $f=t_d/t_r$
  on noise amplitude $\varepsilon$ and reversal 
  time $t_r$ shown by color for $K=3$ (left) and 
  $K=8$ (right).
  The values of $f$ are averaged over $10^6$ 
  trajectories for each color cell, where
  $t_d$ values are obtained as it is shown in Fig.~\ref{fig1}.
}
\end{figure}

The time reversal is
done at time moments $t_r=30$ and $40$
at the middle between two kicks as described above.
For the case at $K=3$ the numerical simulations,
done with double precision
(round-off errors being about $\varepsilon \sim 10^{-16} $),
show the return to the initial state
at time $t=2t_r$ for $t_r=30, 40$
with the energy diffusion restarting for times $t>2t_r$.
However, for $K=8$,
similar to the Chirikov standard map \cite{chirikov1979},
we have the Lyapunov exponent
$\Lambda \approx \ln(K/2) \approx 1.39$
and the exponential error growth
leads to  a large
 accumulated round-off errors $10^{-16} \exp(\Lambda t_r) \sim 100$

\begin{figure}[htbp]
    \centering
    \includegraphics[width=0.48\textwidth]{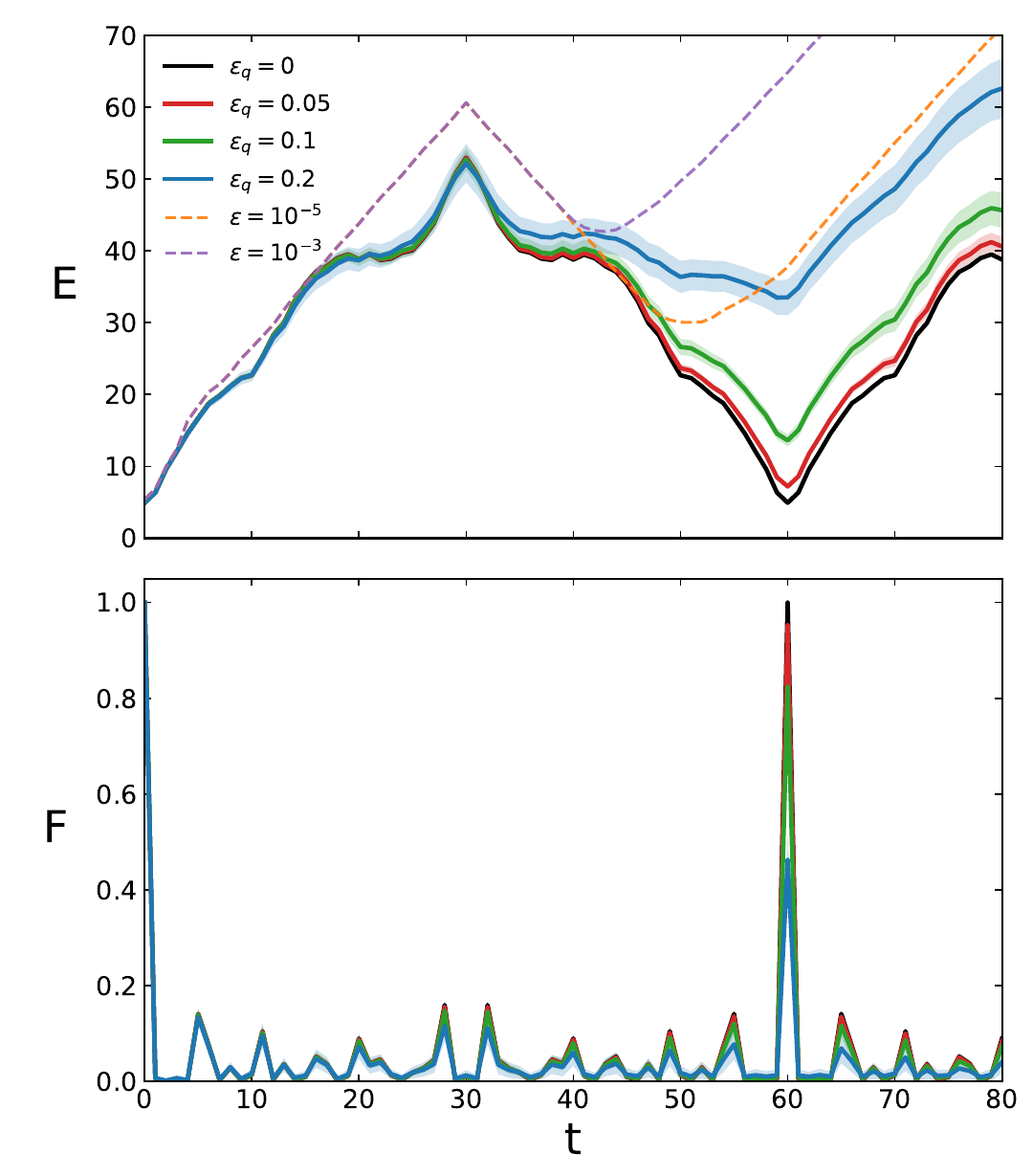}
    \caption{\label{fig3} 
      Time evolution of the mean energy 
      $E(t) = \langle n \rangle$ (top panel) and the quantum fidelity
      $F(t) = \mid \langle \psi(t=0,\varepsilon_q=0) \mid \psi(t,\varepsilon_q) \rangle \mid^2$
      (bottom panel) for $K=3$, with $\hbar=q=1$. Solid curves correspond 
      to different quantum noise amplitudes $\varepsilon_q$: 
      $0$ (black), $0.05$ (red), $0.1$ (green), and $0.2$ (blue). 
      Shaded areas represent the standard deviation computed over 
      $1000$ noise realizations of the quantum map in Eq.~(\ref{eq2}). 
      Dashed curves show the classical mean energy $E(t)$ averaged over 
      $10^6$ trajectories for noise amplitude $\varepsilon=10^{-5}$ (orange)
      and $10^{-3}$ (purple). 
      The initial classical and quantum state distributions, 
      centered at $x_0=\pi, p_0=0$, are shown in the 
      top panels of Fig.~\ref{fig4}.
    }
\end{figure}

To illustrate the effect of errors we introduce
after each time moment $0 \le t \le 2t_r$ 
an additional random variation
of momentum $p_{t+1}  = p_t +\delta_t$
with $-\varepsilon \le \delta_t \le \varepsilon$.
The effects of these artificial noise errors
on energy anti-diffusion are shown in Fig.~\ref{fig1}.
At a given $\varepsilon$ this noise
breaks time reversal and the anti-diffusion
back to the initial state energy
continues only during a finite time $t_d < t_r$.
The dependence of the ratio $f=t_d/t_r$ on $\varepsilon$
is shown in Fig.~\ref{fig2}
for different $t_r$ values.
The results clearly show that the time scale $t_d$
is logarithmically short
$t_d \propto |\ln \varepsilon| /\Lambda$ 
due to exponential growth of errors.
The time evolution of classical density distribution of trajectories
is shown in video files of
Supplementary Material  (SupMat)
for $t_r=30$ and $K=3$, $\varepsilon =0.001$.

{\it Time reversal of quantum chaos.-}
The time reversal of quantum chaos diffusion in (\ref{eq2})
with $t_r=30$ and $K=3$  is shown in Fig.~\ref{fig3} (top panel).
For $t \leq t_r$ there is a diffusive growth of
oscillator energy $E(t)$ being the same as
for the classical case in Fig.~\ref{fig1}.
After the time reversal there is anti-diffusion
back to the initial state during $t_r < t \le 2t_r$ and
for $t>2t_r$ the quantum diffusion restarts again.
For the quantum fidelity 
$F(t) = \mid \langle \psi(t=0,\varepsilon_q=0) \mid \psi(t,\varepsilon_q) \rangle \mid^2$
there is a decrease of $F$
for $0 < t \leq t_r$ followed by its
revival back to $F(2t_r) =1$ for $\varepsilon_q =0$.
In the presence of quantum phase noise $\varepsilon_n$
the time reversal signal is slowly decreasing
with an increase of noise amplitude $\varepsilon_q$
as it is well seen in Fig.~\ref{fig3}.
However, the quantum evolution remains much more
stable in respect to quantum errors
comparing to the case of classical chaotic dynamics
with classical errors. Similar results for $K=8, \hbar=1$ are shown
in SupMat Fig.S1.

\begin{figure}[htbp]
    \centering
    \includegraphics[width=0.48\textwidth]{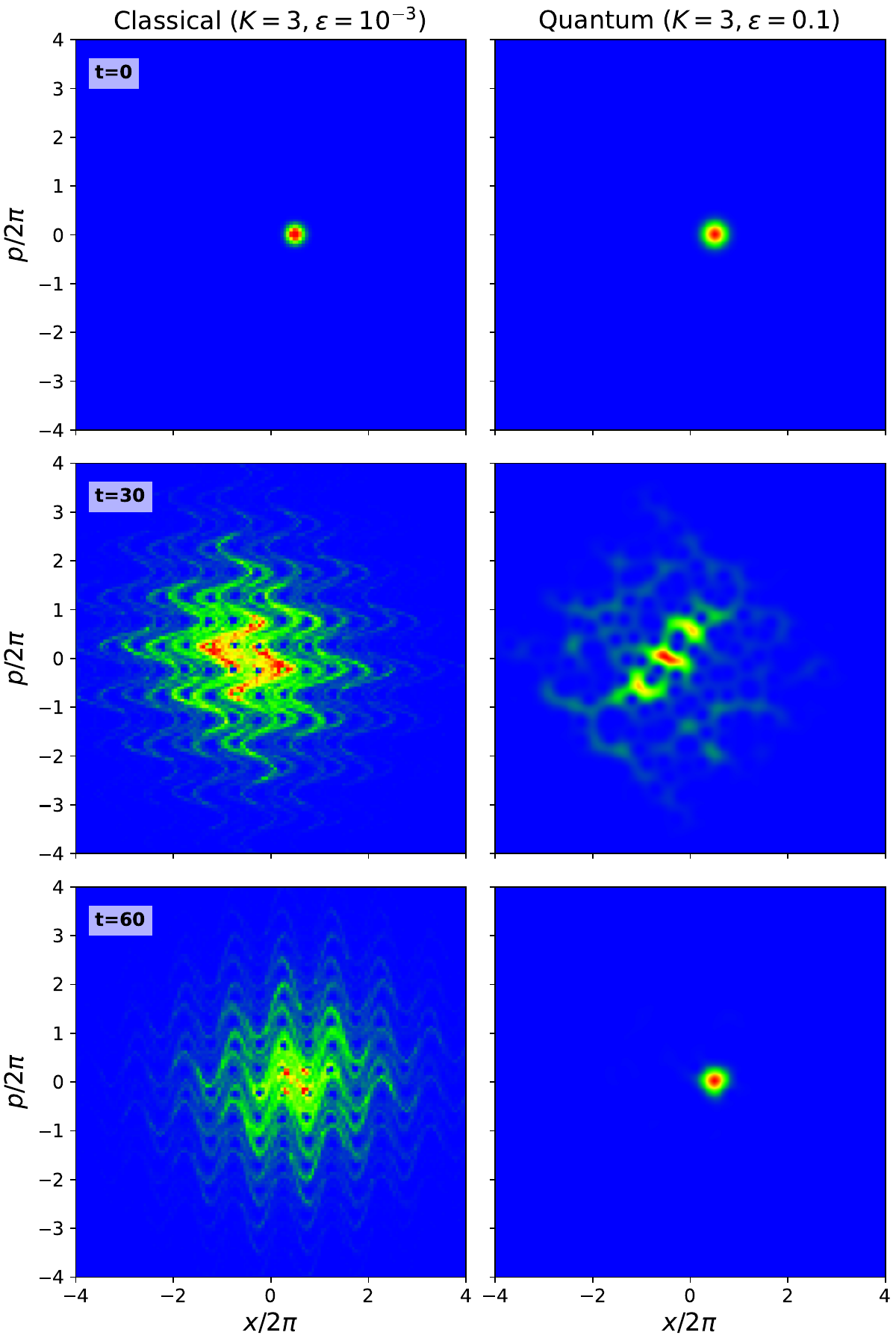}
    \caption{\label{fig4}
      Phase-space comparison of the time-reversal dynamics for classical probability 
      densities (left column) and quantum Husimi distributions \cite{haake,frahm} 
      (right column). The rows, from top to bottom, correspond to the initial state at $t=0$, 
      the state at the reversal time $t=t_r=30$, and the final state at $t=2t_r=60$. 
      Classical distributions are computed from an ensemble of $10^6$ trajectories 
      with a noise amplitude $\varepsilon=10^{-3}$. 
      The quantum Husimi distributions 
      are shown for one noise realization 
      with amplitude $\varepsilon_q=0.1$. 
      All initial states are centered at $(x_0, p_0) = (\pi, 0)$. 
      The system parameters are $K=3$ and $\hbar=1$, 
      with non-linearity exponent $q=1$;
      here red color is for maximal density, blue for zero.
    }
\end{figure}

The examples of classical and quantum distributions
in the phase space $(x,p)$ at time moments
$t=0, t_r, 2t_r$ are shown in Fig.~\ref{fig4}.
At $t=t_r=30$ the classical and quantum distributions
cover  a large area in the phase space.
However, at the return time $t=2t_r=60$
the quantum distribution,
with noise errors amplitude $\varepsilon_q=0.1$,
returns almost perfectly back to the initial state
(with fidelity $F(2t_r) \approx 0.85$).
In contrast, for the classical chaotic dynamics,
with error amplitude $\varepsilon = 0.001$,
the time reversal is broken
and a big fraction of trajectories 
continues to spread diffusely
in the phase space. The videos of these classical 
and quantum evolution
are presented in SupMat.

\begin{figure}[ht]
  \centering
  \includegraphics[width=0.48\textwidth]{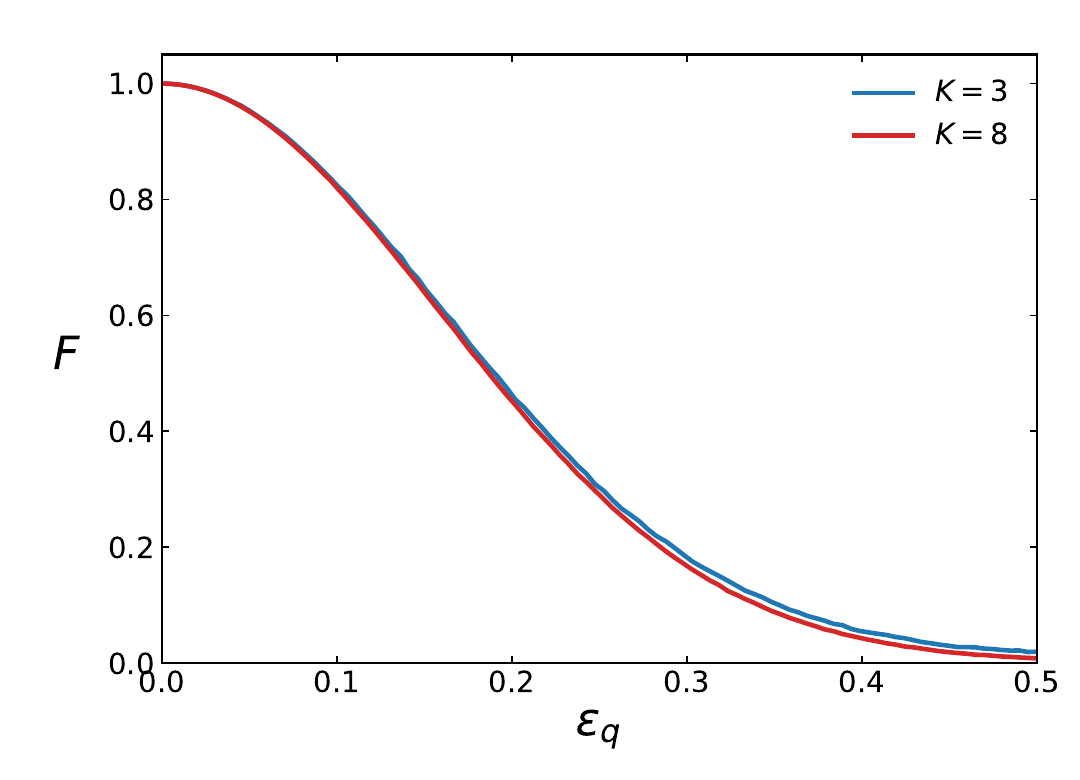}
  \caption{\label{fig5}
    Dependence of the quantum fidelity $F(t=2t_r)$ 
    on the quantum noise amplitude $\varepsilon_q$ 
    for $K=3$ (blue) and $K=8$ (red) 
    at a fixed reversal time $t_r=30$. Values of 
    $F$ are averaged over $1000$ quantum noise realizations.
   }
\end{figure}

\begin{figure}[htbp]
    \centering
    \includegraphics[width=0.48\textwidth]{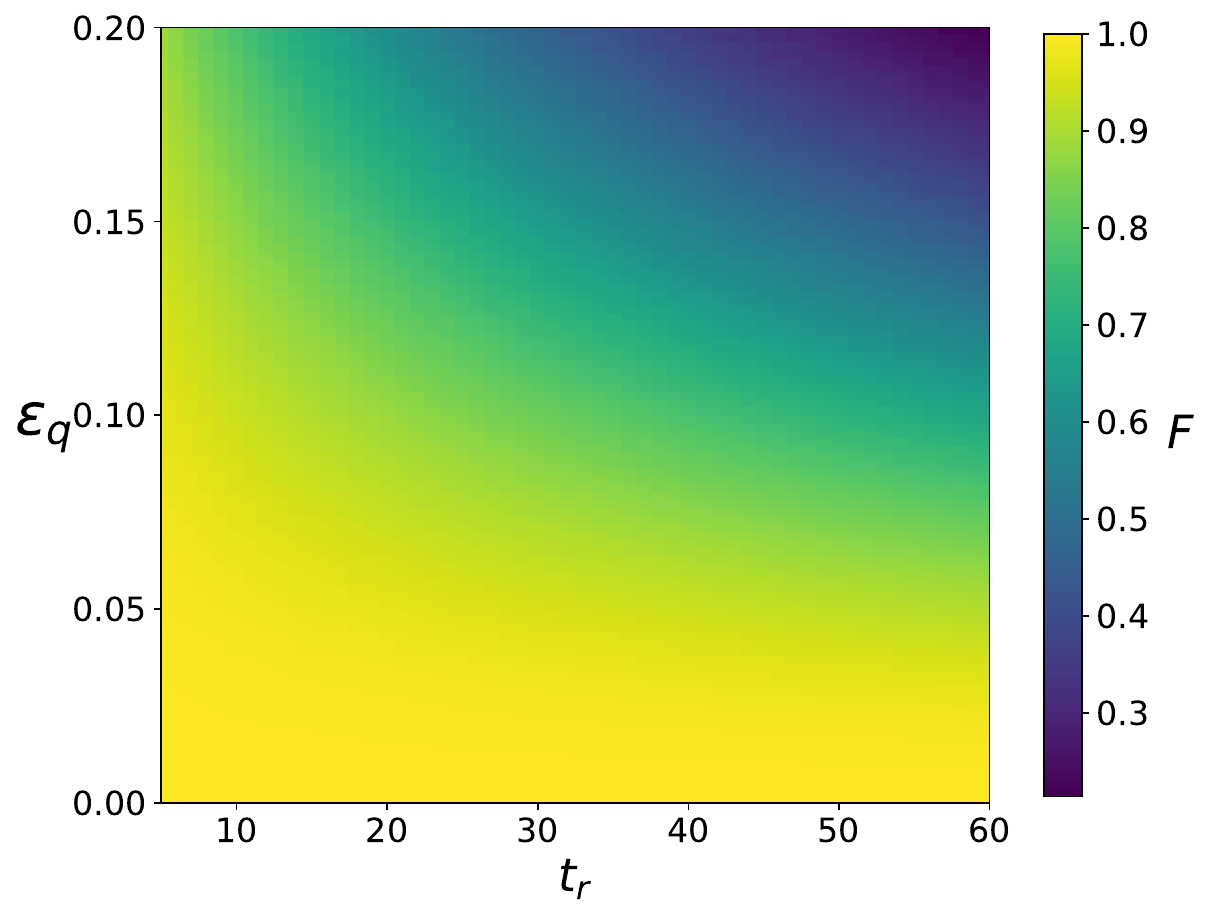}
    \caption{\label{fig6}
      Dependence of the quantum fidelity $F(t=2t_r)$ on 
      the reversal time $t_r$ and the quantum noise 
      amplitude $\varepsilon_q$ for $K=3$, with $\hbar=q=1$. 
      Values of $F$, represented in color scale, are averaged 
      over $1000$ quantum noise realizations.
      }
\end{figure}

In Fig.~\ref{fig5} we show the dependence of
fidelity $F(t=2t_r =60)$ on the
quantum noise amplitude  $\varepsilon_q$. It can be approximately
described by the relation $F \sim \exp(- G\varepsilon^2 t_r)$
where $G$ depends on chaos parameter $K$ and $\hbar$
in agreement with general properties
of Loschmidt echo decay
(see \cite{jalabert2} and Refs. therein).
The results of Fig.~\ref{fig6}
also show that the time reversal fidelity
is very stable in respect to quantum errors 
being in a drastic difference with the exponential sensitivity
of classical dynamics in respect to
classical errors 
shown in Fig.~\ref{fig2}.

We note that the described time reversal procedure works
for noninteracting particles
while their interactions break
reversibility that allows to study
effects of interactions.

 {\it Possible experiments.-} For cold atoms in a kicked optical lattice
a time interval between kicks was about
$T \sim 1\mu s - 30 \mu s $ \cite{raizen,garreau}.
The kick period can be even longer
being comparable  with a trap oscillator period of cold atoms 
with long coherence times. As discussed in \cite{zoller}
the quantum Zaslavsky web map can be also
realized with cold ion traps
with oscillation frequencies of about 1 MHz \cite{blatt}.

{\it Discussion.-} We highlighted the Boltzmann-Loschmidt dispute
on time reversibility of statistical laws \cite{loschmidt,boltzmann2} emerging from
dynamical equations of motion
from the modern view of quantum mechanics
of cold atoms, or ions, in a harmonic trap
and pulsed optical lattice.
We argue that the actual experimental abilities allow to
realize time reversal of quantum chaos diffusion  of cold atoms
with almost 100\% efficiency.

\noindent {\bf Acknowledgments:}
This work has been partially supported through the grant
NANOX $N^o$ ANR-17-EURE-0009 in the framework of 
the Programme Investissements d'Avenir (project MTDINA).

\noindent {\bf Data availability.-} All obtained data are contained in this article.

%%%%%%%%%%%%%%%%%%%%%%%%%%%%%%%%%%%%%%%%%%%%%%%%%%%%%%%%%

\vspace{0.5cm}
\centerline{\bf Supplementary Material}
%\centerline{\bf Boltzmann-Loschmidt dispute reloaded quantum 150 years later}
\vspace{0.3cm}

\setcounter{figure}{0} \renewcommand{\thefigure}{S\arabic{figure}} 
\setcounter{equation}{0} \renewcommand{\theequation}{S.\arabic{equation}} 

The Supplementary Material includes Fig.~S1 and two video files in MP4 format.

For Fig.~S1, the system parameters are similar to those of Fig.~3 in the main text.

\begin{figure}[htbp]
    \centering
    \includegraphics[width=0.48\textwidth]{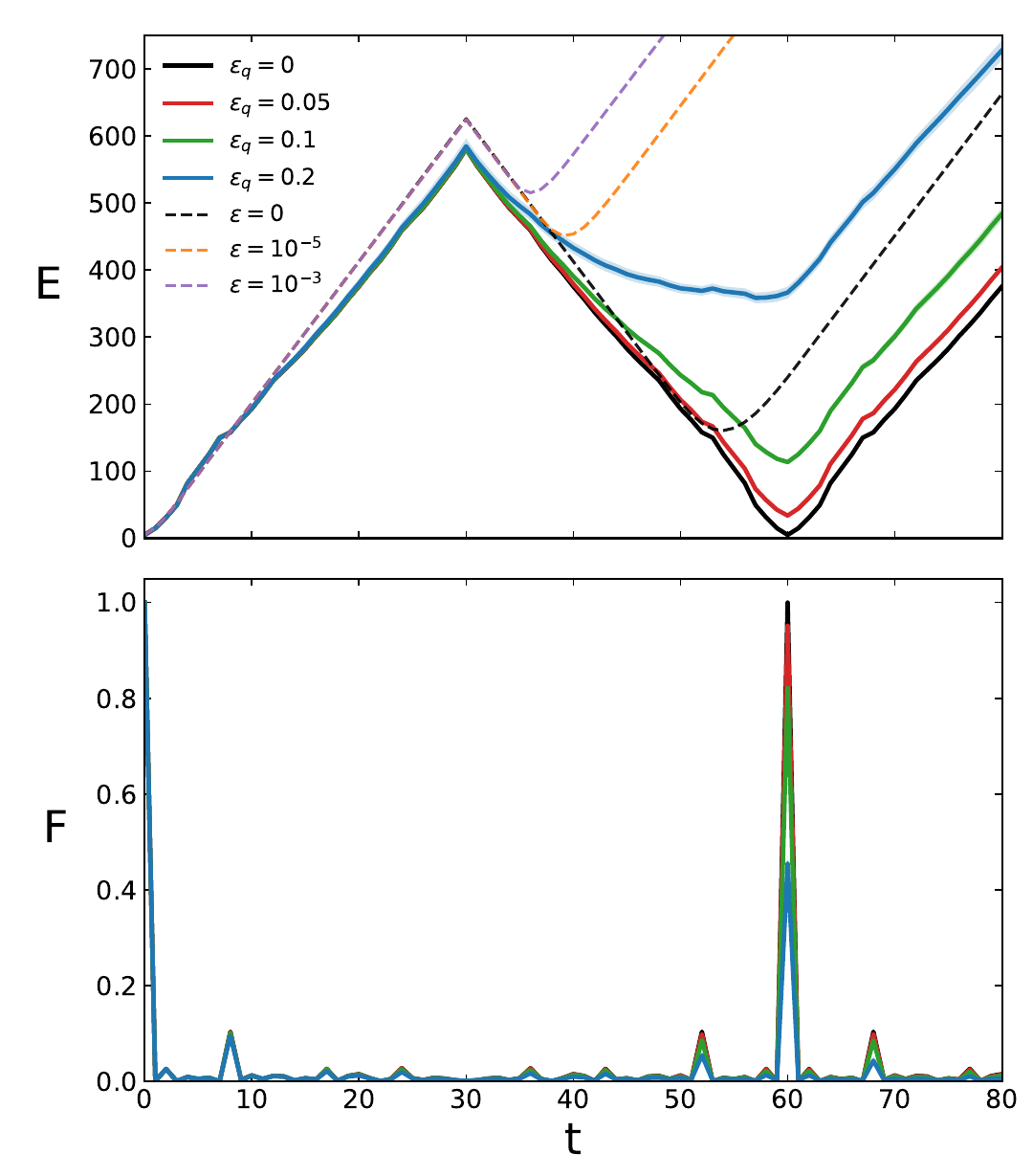}
    \caption{\label{figS1} 
      Time evolution of the mean energy 
      $E(t) = \langle n \rangle$ (top panel) and the quantum fidelity
      $F(t) = \mid \langle \psi(t=0,\varepsilon_q=0) \mid \psi(t,\varepsilon_q) \rangle \mid^2$ (bottom panel) 
      for $K=8$, with $\hbar=q=1$. Solid curves correspond 
      to different quantum noise amplitudes $\varepsilon_q$: 
      $0$ (black), $0.05$ (red), $0.1$ (green), and $0.2$ (blue). 
      Shaded areas represent the standard deviation computed over 
      $100$ noise realizations of the quantum map in Eq.~\ref{eq2}). 
      Dashed curves show the classical mean energy $E(t)$ averaged over 
      $10^6$ trajectories for noise amplitude 
      $\varepsilon=0$ (black); $10^{-5}$ (orange) and $10^{-3}$ (purple). 
      The initial classical and quantum state distributions are
      centered at $x_0=\pi, p_0=0$.
    }
\end{figure}

The video file \texttt{supmatvideo1.mp4} shows
the classical time evolution of the density distribution in the phase space $(x,p)$
(bottom panel). The top panel shows the dependence of the classical energy $E(t)$
and classical fidelity $F(t)$ at a noise amplitude of $\varepsilon=10^{-3}$
obtained with $10^6$ classical trajectories for one noise realization
and the parameters of Fig.~4 ($K=3,  q=1$).
The classical fidelity $F(t)$ is defined as an overlap
between the initial classical phase-space distribution 
at $t=0$ and the distribution at time $t$:
\begin{equation}
F(t) = \frac{\int \rho(x, p, 0) \rho(x, p, t) \, dx dp}
{\sqrt{\int \rho^2(x, p, 0) \, dx dp \int \rho^2(x, p, t) \, dx dp}},
\end{equation}
where $\rho$ denotes the classical coarse-grained density distributions.

The video file \texttt{supmatvideo2.mp4} shows 
the quantum time evolution of the Husimi density distribution
in the phase space $(x,p)$
(bottom panel) for one noise realisation.
The top panel shows the dependence of the quantum energy $E(t)$
and quantum fidelity $F(t)$ at a noise amplitude of $\varepsilon_q=0.1$
for one noise realization and the parameters of Fig.~4 ($K=3, \hbar=1, q=1$).


\begin{thebibliography}{99}
\bibitem{boltzmann1} L.~Boltzmann, {\it Weitere Studien \"uber 
        das W\"arme\-gleich\-gewicht unter Gasmolek\"ulen},
        Wiener Berichte {\bf 66}, 275 (1872).
\bibitem{loschmidt} J. Loschmidt, {\it \"Uber den Zustand des W\"arme\-gleich\-gewichts eines
  Systems von K\"orpern mit R\"ucksicht auf die Schwerkraft},
  Sitzungsberichte der Akademie der Wissenschaften, Wien {\bf II 73},  128 (1876).
\bibitem{boltzmann2} L. Boltzmann, {\it \"Uber die Beziehung eines allgemeine
  mechanischen Satzes zum zweiten Haupsatze der W\"armetheorie},
  Sitzungsberichte der Akademie der Wissenschaften, Wien {\bf II 75},  67 (1877).
\bibitem{mayer}  J.E. Mayer, M. Goeppert-Mayer, {\it Statistical mechanics},
       John Wiley \& Sons, N.Y. (1977).
\bibitem{arnold} V.~Arnold, A.~Avez, {\it Ergodic problems of classical mechanics},
       Benjamin, N.Y. (1968).
\bibitem{sinai} I.~P.~Cornfeld, S.~V.~Fomin and Ya.~G.~Sinai,
        {\it Ergodic theory}, Springer-Verlag, N.Y. (1982).
\bibitem{chirikov1979} B.~V.~Chirikov,
  {\it A universal instability of many-dimensional oscillator systems},
  Phys. Rep. {\bf 52}, 263 (1979).
\bibitem{lichtenberg}A.~Lichtenberg and M.~Lieberman, 
  {\em Regular and Chaotic Dynamics}, Springer, N.Y. (1992).
\bibitem{dls1983} D.~L.~Shepelyansky, {\it Some statistical properties of simple
  classically stochastic quantum systems}, Physica D {\bf 8}, 208 (1983).
\bibitem{sklar} L.~Sklar, {\it Physics and chance: phylosophical issues in the foundations
  of statistical mechanics}, Cambridge Univ. Press, UK (1993).
\bibitem{lindey} D.~Lindey, {\it Boltzman's atom: the great debate that
  launched a revolution in physics}, The Free Press, New York (2001).
\bibitem{bader} A.Bader, and L.~Parker,
  {\it Joseph Loschmidt, Physicist and Chemist},
  Physics Today {\bf 54(3)}, 45 (2001).
\bibitem{price} H.~Price, {\it Boltzmann's Time Bomb},
  The British Journal for the Philosiphy of Science,
  {\bf 53(1)}, 83 (2002);
  \url{https://doi.org/10.1093/bjps/53.1.83}.
\bibitem{uffink} J.~Uffink, {\it Compendium of the foundations of classical statistical physics},
  philsci-archive.pitt.edu (2006);
  \url{https://philsci-archive.pitt.edu/2691/1/UffinkFinal.pdf}
  (Accessed  January 23, 2026).
\bibitem{uffink1} J.~Uffink, {\it Boltzmann’s Work in Statistical Physics},
  Stanford Encyclopedia of Philosophy Archive (2024);
  \url{https://plato.stanford.edu/archives/win2024/entries/statphys-Boltzmann/}
   (Accessed  January 23, 2026).
\bibitem{weidenmuller} H.A.~Weidenmuller,
  {\it The rise of stochasticity in physics},
  Eur. Phys. J. Plus {\bf 140}, 304 (2025).
\bibitem{ehrenfest} P.~Ehrenfest,
  {\it Bemerkung über die angenaherte Gultigkeit der klassischen Mechanik
    innerhalb der Quantenmechanik}, Zeitschrift fur Physik {\bf 45}, 455 (1927).
\bibitem{chirikov1981} B.V.~Chirikov, F.M.~Izrailev, and  D.L.~Shepelyansky,
  {\it Dynamical stochasticity in classical and quantum mechanics},
  Sov. Scient. Rev. C {\bf 2}, 209 (1981).
\bibitem{chirikov1986}  B.V.~Chirikov, F.M.~Izrailev, and  D.L.~Shepelyansky,
  {\it Quantum chaos: localization vs. ergodicity}, Physica D {\bf 33}, 77 (1988).
\bibitem{haake} F.~Haake,
  {\it Quantum signatures of chaos}, Springer, Berlin (2001). 
\bibitem{dls1981} D.L.~Shepelyanskii, {\it Dynamical stochasticity in nonlinear quantum systems},
  Theor. Math. Phys. {\bf 49(1)}, 925 (1981).
\bibitem{dlsehrenf} D,Shepelyansky, {\it Ehrenfest time and chaos},
  Scholarpedia {\bf 15(9)}, 55031 (2020).
  % doi:10.4249/scholarpedia.55031
\bibitem{peres} A.~Peres, {\it  Stability of quantum motion in chaotic and regular systems},
  Phys. Rev. A {\bf 30}, 1610 (1984).
\bibitem{jalabert1} R.A.~Jalabert, and H.M.~Pastawski,
  {\it The semiclassical tool in complex physical systems: Mesoscopics and decoherence}
  Adv.  Solid State Phys., {\bf 41}, 483 (2001).
\bibitem{frahm} K.M.~Frahm, R.~Fleckinger, and D.L.~Shepelyansky,
  {\it Quantum chaos and random matrix theory for fidelity decay
    in quantum computations with static imperfections},
   Eur. Phys. J. D {\bf 29}, 139 (2004).
\bibitem{prosen} T.~Gorin, T.~Prosen, T.H.~Seligman, and M.~Znidaric,
  {\it Dynamics of Loschmidt echos and fielity decay}. Phys. Rep., {\bf 435},  33 (2006). 
\bibitem{jacquod} P.~Jacquod, and C.~Petitjean,
  {\it Decoherence, entanglement and irreversibility in quantum dynamical systems
    with few degrees of freedom}. Adv.  Phys. {\bf 58},  67 (2009).
\bibitem{jalabert2} A.~Gousev, R.A.~Jalabert, H.M~Pastavsli, and D.A.~Wisniacki,
  {\it Loschmidt echo}, Scholarpedia {\bf 7(8)}, 11687 (2012).
  % doi:10.4249/scholarpedia.11687
\bibitem{hahn1} E.L.~Hahn, {\it Spin echos},
  Phys. Rev. {\bf 80}, 580 (1950).
\bibitem{hahn2} {\it Pulsed Magnetic Resonance: NMR, ESR, and Optics: A
       Recognition of E. L. Hahn}, edited by D. M. S. Bagguley
       (Oxford University Press, New York, 1992).
\bibitem{pastaw} H.M.~Pastawski, P.R.~Levstein, G.~Usaj, J.~Raya, and J.~Hirschinger,
  {\it A nuclear magnetic resonance answer to the Boltzmann-Loschmidt controversy?},
  Physica A {\bf 283}, 166 (2000).
%\bibitem{fink1} M.~Fink, D.~Cassereau, A.~Derode, C.~Prada, P.~Roux, M.~Tanter,
%  J.-L.~Thomas, and F.~Wu, {\it Time-reversed acoustics},
%  Rep. Prog. Phys. {\bf 63}, 1933 (2000).
\bibitem{fink1} M.~Fink, {\it Chaos and time-reversed acoustics},
  Phys. Scripta {\bf 2001(T90)}, 268 (2001).
\bibitem{fink2} G.~Lerosey, J. de Rosny, A.~Tourin, A.~Derode, G.~Montaldo, and M.~Fink,
  {\it Time Reversal of Electromagnetic Waves}, Phys. Rev. Lett. {\bf 92}, 193904 (2004).
\bibitem{geopht} C.S.~Larmat, R.A.~Guyer, and P.A.~Johnson,
  {\it Time-reversal methods in geophysics},
  Phys. Today {\bf 63(8)}, 31 (2010).
\bibitem{qbl1}  J.~Martin, B.~Georgeot and D.L.~Shepelyansky,
  {\it Cooling by time reversal of atomic matter waves},
  Phys. Rev. Lett. {\bf 100}, 044106 (2008).
\bibitem{qbl2}  J.~Martin, B.~Georgeot and D.L.~Shepelyansky,
  {\it Time reversal of Bose-Einstein condensates},
  Phys. Rev. Lett. {\bf 101}, 074102 (2008).
\bibitem{bec1} A.~Ullah, and M.D.~Hoogerland,
  {\it Experimental observation of Loschmidt time reversal of a quantum chaotic system},
  Phys. Rev. E {\bf 83}, 046218 (2012). 
\bibitem{bec2} A.~Cao, R.~Sajjad, H.~Mas, E.Q.~Simmons, J.L.~Tanlimco, E.~Nolasco-Martinez,
  T.~Shimasaki, H.E.~Kondakci, V.~Galitski, and D.~Weld.
  {\it Interaction-driven breakdown of dynamical localization in a kicked quantum gas},
  Nature Phys. {\bf 18}, 1302 (2022). 
\bibitem{zaslav1} G.M.~Zaslavskii, M.Y.~Zakharov, R.Z.~Sagdeev, D.A.~Usikov, and A.A.~Chernikov,
  {\it Stochastic web and diffusion of particles in magnetic field},
  Sov. Phys. JETP {\bf 64}, 294 (1986).
\bibitem{zaslav2} A.A.~Chernikov, R.Z.~Sagdeev, D.A.~Usikov, A.Y.~Zakharov, and
  G.M.~Zaslavsky {\it Minimal chaos and stochastic web},
  Nature  {\bf 326(9)}, 559 )1987).
\bibitem{zaslavweb} G.~Zaslavsky, {\it Zaslavsky web map},
  Scholarpedia  {\bf 2(10)}, 3369 (2007).
\bibitem{webspace} The Zaslavsky web map generator,
  \url{https://www.russellcottrell.com/fractalsEtc/Zaslavsky.htm}
  (Accessed February 2, 2026).
\bibitem{sire} D.~Shepelyansky, and C.~Sire,
  {\it Quantum evolution in a dynamical quasi-crystal},
  Europhys, Lett.  {\bf 20(2)}, 95 (1992).
\bibitem{dana} I.~Dana, {\it Quantum suppression of diffusion on stochastic web},
  Phys. Rev. Lett.  {\bf 73}, 1609 (1994).
\bibitem{zoller} S.A~Gardiner, J.I.~Cirac, and P.~Zoller,
  {\it Quantum chaos in an ion trap: the delta-kicked harmonic oscillator},
  Phys. Rev. Lett.  {\bf 79}, 4790 (1997).
\bibitem{kells} G.A.~Kells, J.~Twamley, and D.M.~Heffernan,
  {\it Dynamical properties of the delta-kicked harmonic oscillator},
  Phys. Rev. E {\bf 79}, 015203(R) (2004).
\bibitem{buchleitner} A.R.R.~Carvalho, and A.~Buchleitner,
  {\it Web-assisted tunneling in the kicked harmonic oscillator},
  Phys. Rev. Lett.  {\bf 93}, 204101 (2004).
\bibitem{gardiner} T.P.~Billam, and S.A.~Gardiner,
  {\it Quantum resonances in an atom-optical $\delta$-kicked harmonic oscillator},
  Phys. Rev. A  {\bf 80}, 023414 (2009).
\bibitem{raizen} F.L.~Moore, J.C.~Robinson, C.F.~Bharucha, B.~Sundaram, and M.G.~Raizen,
  {\it Atom optics realization of the quantum $\delta$-kicked rotor},
  Phys. Rev. Lett. {\bf 75}, 4598 (1995).
\bibitem{lima} R.~Lima, and D.~Shepelyansky,
  {\it Fast delocalization in a model of quantum kicked rotator},
  Phys. Rev. Lett.  {\bf 67}, 1377 (1991).
\bibitem{artuso} R.~Artuso,  {\it Kicked Harper model},
  Scholarpedia  {\rm 6(10)}, 10462 (2011).
\bibitem{flux} A.D.~Chepelianskii, and D.L.~Shepelyansky,
  {\it Kicked fluxonium with quantum strange attractor},
  MDPI Physics {\bf 8}, 22 (2026).
\bibitem{garreau} J.~Chabe, G.~Lemarie, B.~Gremaud, D.~Delande, P.~Sziftgiser,
  and J.C.~Garreau, {\it Experimental Observation of the Anderson Metal-Insulator Transition
  with Atomic Matter Waves}, Phys. Rev. Lett. {\bf 101}, 255702 (2008).
\bibitem{blatt} I.~Pogorelov,  T.~Feldker, Ch.D.~Marciniak ,  L.~Postler,  G.~Jacob,
  O.~Krieglsteiner, V.~Podlesnic, M.~Meth, V.~Negnevitsky, M.~Stadler ,  B.~Höfer,
  C. Wachter,  K.~Lakhmanskiy, R.~Blatt, P.~Schindler,  and T.~Monz,
  {\it Compact Ion-Trap Quantum Computing Demonstrator}, PRX Quantum {\bf 2}, 020343 (2021).
  
\end{thebibliography}
\end{document}